# The CMS Level-1 Trigger at LHC and Super-LHC

C. Foudas
*Imperial College, Prince Consort Road, London, SW7 2BW, UK*

The Level-1 trigger of the CMS experiment at CERN has been designed to select proton-proton interactions whose final state includes signatures of new physics in the form of high transverse energy electrons, photons, jets, or high missing transverse energy. The Level-1 trigger system process data in a pipeline fashion at a rate of 40 MHz, has a design latency of 128 bunch crossings and an output rate of 100 KHz. The design of this system is presented with emphasis on the calorimeter triggers. After a long period of testing and validation of its performance the Level-1 trigger system has been installed and commissioned at the CMS experiment at CERN. Cosmic ray data and Monte Carlo events have been used to compare the actual performance of the trigger with expectations from off-line emulation models. Results from these studies are presented here. The limitations of this system to cope with future luminosity upgrades of the LHC, the Super-LHC, are discussed. The current CMS plan for a new CMS Level-1 trigger system at the Super-LHC is presented. The center point of the new system is a Level-1 tracking trigger which uses data from a new CMS silicon tracking detector.

## 1. THE CMS TRIGGER SYSTEM

The running conditions at LHC impose severe constraints on the design of trigger systems mainly due to the enormous data rate and the harsh radiation environment. At an instantaneous luminosity of $10^{34}$ cm$^{-2}$ sec$^{-1}$, an average of 22 interactions (minimum bias events) are expected to occur in each beam crossing possibly superimposed with interactions originating from new physics. The large number of minimum bias events per crossing combined with small cross-sections of possible discovery signatures requires a sophisticated online event selection system. This is achieved by the CMS trigger and DAQ system in two stages: The Level 1 trigger (Lvl-1) and the Higher Level Trigger system (HLT) [1].

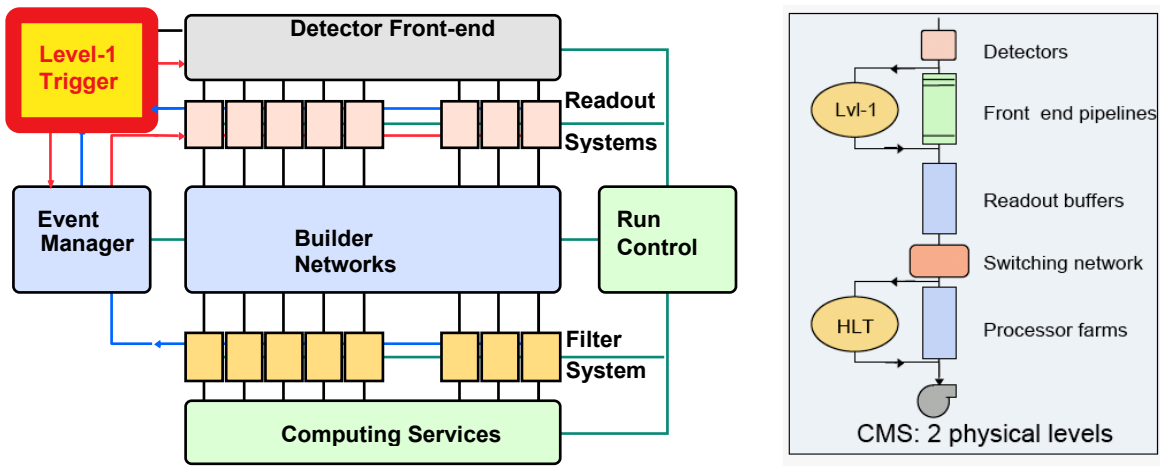

Figure 1: The CMS data acquisition system (left). The Level-1 trigger decision is distributed to the detector front end (top) as well as to off detector readout systems. Builder networks using cross-point switches construct the event record from the event fragments which come from different detector parts. The HLT (Filter System) receives and processes complete events. The CMS Level-1 trigger (right) receives data from the calorimeter and the muon detectors and produces a yes/no decision a fixed number of crossings later (latency). Pipeline memories of depth equal to the trigger latency store the data on the detector until the Level-1 decision arrives.

The CMS trigger and DAQ system is shown in Fig. 1. Data from calorimeters and the muon system are used by the Level-1 trigger (Lvl-1) to perform the first level of online event selection using custom electronic systems. The algorithms process data in pipeline fashion, using pattern recognition and fast summing techniques, without introducing dead-time. The algorithms use input data of reduced granularity and resolution. The calorimeters transmit energy sums along with data that carry information to characterise the energy deposition (electron/photon, tau, narrow jet, muon). Similarly in the muon system the hits are summarized in terms of track stubs. The Lvl-1 output quantities are muon, electron, photon, tau and jet candidates along with jet counts and global transverse and missing transverse energy sums. The decision is derived by the CMS Global Trigger (GT) which imposes cuts on the Lvl-1 output quantities.

The current CMS Lvl-1 is segmented in terms of detectors. Each detector has a Trigger Primitive Generator (TPG) which provides condensed data, followed by a Regional Trigger where the first level of trigger algorithms is executed. At the next level, the Global Calorimeter Trigger (GCT) and the Global Muon Trigger (GMT) execute the final detector-specific trigger algorithms. Finally, data from the GCT and GMT are transmitted to GT where the Lvl-1 Accept (L1A) decision is made.

In this scheme physics objects from different detectors, for example muon candidates or missing transverse energy from the calorimeters, can only be combined after arrival at the GT. Hence, in the current scheme there is no possibility of combining information from different detectors to improve the performance of the Lvl-1 physics object reconstruction.

The GT decision arrives at the detector front ends 144 beam crossings (3.6 μsec latency)[1] after the interaction, at a rate which is required to be less than 100 kHz (Poisson average). Triggering at the first level is the most challenging part of the online data selection since it requires very fast custom designed electronics with a significant portion placed on the CMS detector. This introduces severe constraints in the front-end electronics design, which must be radiation tolerant with limited power consumption. Lepton and jet finding algorithms run on large off-detector processors whose hardware is based on a mixture of discrete devices and FPGAs (Field Programmable Gate Arrays).

The HLT software runs on a large computer farm of fast commercial processors. The algorithms have access to data from all CMS sub-detectors, including the tracker, with full granularity and resolution. The HLT reconstruction software is similar to what will eventually be used offline for CMS data analysis. Hence, the HLT algorithms, in contrast with the Lvl-1, calculate quantities with a resolution comparable to the final detector resolution. For example tracking information from pixel and silicon strip detectors is combined with patterns of calorimeter energy deposition to define electron candidates. Tau-jet candidates found in the calorimeter are combined with high $p_T$ stubs in the tracker to form a hadronic tau trigger which has excellent efficiency and purity. These techniques enable the HLT to define its output objects very precisely and significantly reduce background.

The HLT output quantities are similar to those of Lvl-1 but with far better resolution, purity and efficiency. The maximum HLT input rate is 100 KHz and the output rate is about 100 Hz.

---

[1] The current number is larger than the design latency of 128 bunch crossings but could decrease after optimizing the performance of the firmware used for trigger algorithms and data transmission.

Detailed simulations have been performed using a software model of the entire CMS trigger system and its performance is now understood in great detail. This is required to predict expected data rates but also, most important, to understand the efficiency and purity for selecting interactions with discovery potential.

## 2. DATA VALIDATION AND COMMISSIONING OF THE CMS TRIGGER SYSTEM

The CMS trigger system has been installed at the CMS underground cavern, USC-55 in 2007. This system has undergone a series of detailed tests which tested data transmission and synchronization as well as the validity of the trigger algorithms. A summary of these tests is given in this section.

### 2.1. Synchronization and quality of data transmission tests

The CMS Lvl-1 trigger allows the injection and capturing of test pattern data in various stages of the trigger pipeline. This permits two types of tests. The trigger pipelines were first tested for data integrity to answer the question if an electron jet or muon candidate of a given transverse energy above threshold would cause a trigger. Electron, muon and jet patterns were injected in different stages of the trigger pipeline and the proper data transmission was established. Subsequently, different trigger candidates were injected in different bunch crossings in the trigger pipeline and this established also that the trigger system was properly synchronized and would produce triggers at the correct bunch crossing. Both tests established that the off detector trigger electronics were performing correctly as intended.

Next the trigger system was connected to the CMS calorimeter and muon detectors and data were collected by triggering on cosmic ray muons. During this time the majority of noisy channels were removed. The fraction of noisy channels was far below 1% of the channels of the relevant detectors. Shown in Fig. 2 are data from the CMS calorimeter trigger collected by triggering on single energy depositions in the CMS hadronic calorimeter. Transverse energy depositions in 4 by 4 trigger tower regions[2], the input to the CMS Lvl-1 jet finder, are shown in Fig. 2 (left) and data corresponding to the same energy depositions recorded by the electron trigger are shown in Fig. 2 (right). The total electron energy is defined as the sum of the energies deposited at the electromagnetic (ECAL) and hadronic (HCAL) calorimeters. For this test the electromagnetic calorimeter depositions have been ignored. In both jet and electron trigger data we see no anomalous energy depositions from noisy channels or malfunctioning electronic channels. Further more both triggers record approximately the same energy depositions within resolution as expected. Similar studies were performed for all triggers and these tests established the functionality of the entire trigger chain.

### 2.2. Trigger data validation using the Lvl-1 trigger emulator

A software package was developed to upload entire Monte Carlo events corresponding to Higgs, SUSY and other exotic signals directly at the inputs of the trigger pipelines. These events were then processed by the trigger hardware and the result was captured at the output of the trigger pipeline. The same events were processed by the off line Lvl-1 trigger emulator software, which is used to calculate trigger efficiencies off line, and was compared with the results from the trigger hardware. Shown in Fig. 3 is such a comparison for Higgs events decaying into 2 electron-positron pairs as seen by the CMS calorimeter isolated electron trigger. The results from processing of Monte Carlo events by the CMS trigger are identical with those from the CMS trigger emulator and this demonstrates that the proper functionality of the electron trigger. The disagreement in the fist transverse energy bin is due to empty events which were mixed with Monte Carlo events as part of this test. Jet triggers were also subjected to the same test before being commissioned.

---

[2] A trigger tower is covers approximately 10×10 cm$^2$ of the detector in the central rapidity region.

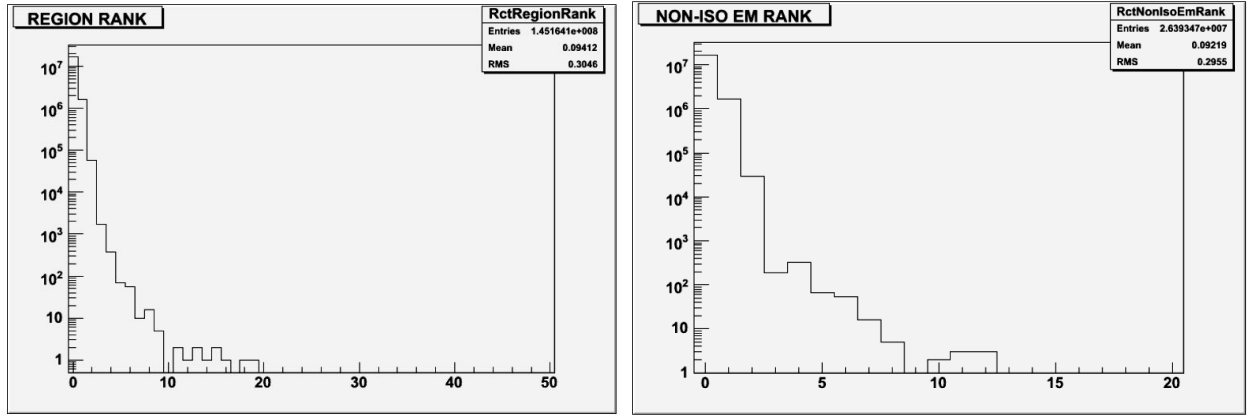

Figure 2: Trigger data from energy depositions in the CMS hadronic calorimeter originating from cosmic ray muons and noise: The regional transverse energy (left) and the total electron energy (right). The x-axis represents energy in arbitrary units which is approximately 1 GeV/count.

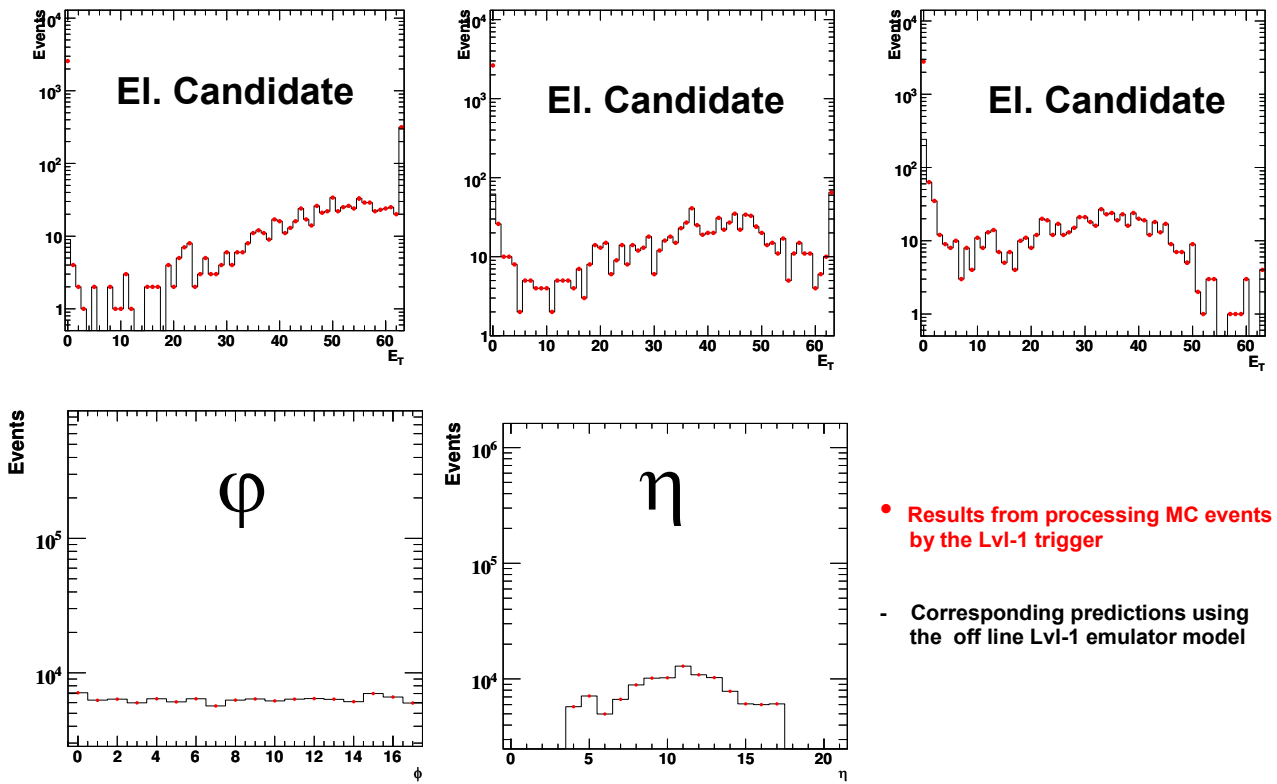

Figure 3: Predictions of the Lvl-1 trigger emulator versus results from the Lvl-1 trigger electron finder for Higgs events decaying into two electron-positron pairs. Transverse energy distributions (top) for the three highest transverse energy electrons detected by the Lvl-1 trigger and the corresponding phi and pseudorapidity distributions (bottom).

## 3. THE CMS TRIGGER AT THE SUPER-LHC

The LHC machine will be upgraded to increase the luminosity by an order of magnitude to $10^{35}$ cm$^{-2}$.s$^{-1}$. The date for the upgrade is not yet defined but is expected to be about 10 years after LHC starts operating and it is referred to as the Super-LHC (SLHC) [2]. The machine parameters are not yet frozen but the clock speed is expected to be 20MHz instead of the current 40MHz.

The impact of detector performance on physics measurements at SLHC depends on the channel considered. Searches at the high-mass frontier, in most cases detection of multi-TeV objects, will not be impaired by the high luminosity environment. In contrast, accurate measurements of objects with mass below 1 TeV will be significantly affected by the large number of minimum bias events, and there is a danger that reduced efficiencies or increased backgrounds could spoil the advantage of the higher luminosity.

The Lvl-1 trigger efficiency and purity to select Higgs and other exotic signals is certain to degrade due to the increase of minimum bias events. It will be a trigger challenge at the SLHC to distinguish a low mass Higgs among hundreds of minimum bias events. The efficiency of all high $p_T$ isolated lepton triggers is certain to decrease due to the strict isolation criteria they employ and because thresholds would have to be raised. Raising the Lvl-1 thresholds reduces the potential of the SLHC by risking to put low mass LHC discoveries (below 1 TeV) out of reach. Jet triggers must also apply higher thresholds to survive in their current state at the SLHC.

In some cases raising thresholds will not achieve sufficient rate reduction. An example is performance of muon Lvl-1 triggers, which suffer from poor transverse momentum resolution at high $p_T$. This is due to multiple Coulomb scattering in the iron between chamber stations, and the limited lever arm for muon tracking. The limited momentum resolution and miss-association with low momentum muons result in a relatively flat high $p_T$ spectrum. Hence the required Lvl-1 rate reduction cannot be achieved by simply increasing $p_T$ thresholds.

In the case of electromagnetic triggers at SLHC, most ECAL signals are generated by photons from $\pi^0$ decays and the secondary particle fluxes will be considerably greater. The shower isolation criteria in the calorimeter used in the L1 trigger cannot be much improved using data only from the ECAL. Therefore a strategy similar to what is presently done in the HLT must be defined, which requires tracking information.

Discoveries at the LHC in the mass range below 1 TeV will require detailed study at the SLHC and thus the trigger efficiency for such signals must be maintained. Gains in background reduction can only come by improving L1T algorithms to achieve higher purity for selecting physics signals. Hence, increased luminosity requires a significant revision of existing trigger strategies. This can only be achieved by using new information in the Lvl-1decision.

The current Lvl-1 design already uses information from all detectors except the tracker, so improvements in background reduction efficiency and purity are expected from using silicon strip and pixel detector information. Hence, designing a new silicon tracker and a Lvl-1 tracking trigger is a high priority for CMS. The two are evidently tightly coupled.

### 3.1. Potential gain from a Lvl-1 tracking trigger at the SLHC

Over the past 20 years, Lvl-1 trigger systems have dramatically improved performance in speed and algorithm complexity by using state of the art technology. Advances such as large Field Programmable Gate Arrays (FPGAs) instrumented with fast data links have provided platforms for sophisticated Lvl-1 trigger processing. Jet triggers operating at Level-1 exemplify algorithms which were typically run in the past during offline event selection or at the HLT level using commercial CPUs.

Thus algorithms, whose performance had been proven at the HLT level, could be implemented at lower trigger levels by following industry trends in telecom and digital processing techniques. This resulted in large data reductions early in the trigger chain as well as Lvl-1 trigger output samples of unprecedented efficiency and purity. This combination also reduces R&D costs and decreases R&D time.

Hence, CMS HLT algorithms [1] are used as a starting point to evaluate the use of tracking in the Lvl-1 trigger decision. Estimates of background rejection can be obtained from studies of algorithms which use tracker data in coincidence with calorimeter and muon detector data to select events:

- The isolated electron trigger suppresses backgrounds by a factor 10 by putting in coincidence calorimeter electron candidates with hits found in the pixel detector. A further factor 3 is gained by including track stubs close to the calorimeter resulting in a factor 30 in background rejection.
- The τ-trigger suppresses backgrounds by a factor 10 when tracks in the pixel detector are put in coincidence with jets found in the calorimeter.
- Muon triggers benefit by a factor of 50 when tracker information is put in coincidence.

The conclusion from all these studies is that: Using tracking information in Lvl-1 trigger algorithms has the potential of compensating for the increase of the backgrounds. A large benefit is obtained by using tracks, hits and vertices found by the pixel detector which in turn creates severe challenges for any tracking trigger due to the stringent constraints in terms of power, radiation hardness and data volume and speed. Advanced information from other trigger systems such as the calorimeter or muon system is essential in reducing the number of hit combinations in the inner tracker. While not all of these are certain to be implementable for SLHC, the benefits are obvious and development of a tracking trigger for the future is a CMS priority. We conclude that R&D must focus on the design and implementation of a new system which could exploit the tracker in the Lvl-1 trigger.

### 3.2. The Stacked Tracking Detector

A possible scheme for triggering has been studied by some of us. The occupancy of a tracking detector is a key point. It has been shown that at 20 MHz beam crossing rate and a luminosity of $10^{35}$ cm$^{-2}$s$^{-1}$ a detector positioned at radius of 10 cm away from the beam axis will be exposed to ~16 hits/cm$^2$/bc [3] which corresponds to a data rate of approximately 20-40 Gb/sec/cm² [3]. This is dominated by low momentum (below 1GeV) particles which curl around the beam axis in the 4T magnetic field. The resulting high data rate presents several obvious problems:

- Transmitting all tracker data out of the detector is completely unrealistic.
- The number of hit combinations to produce track stubs is enormous even if one can initiate track searches from calorimeter objects.
- The on-detector power required by such a large number of fast optical links is far too high for a realistic design.

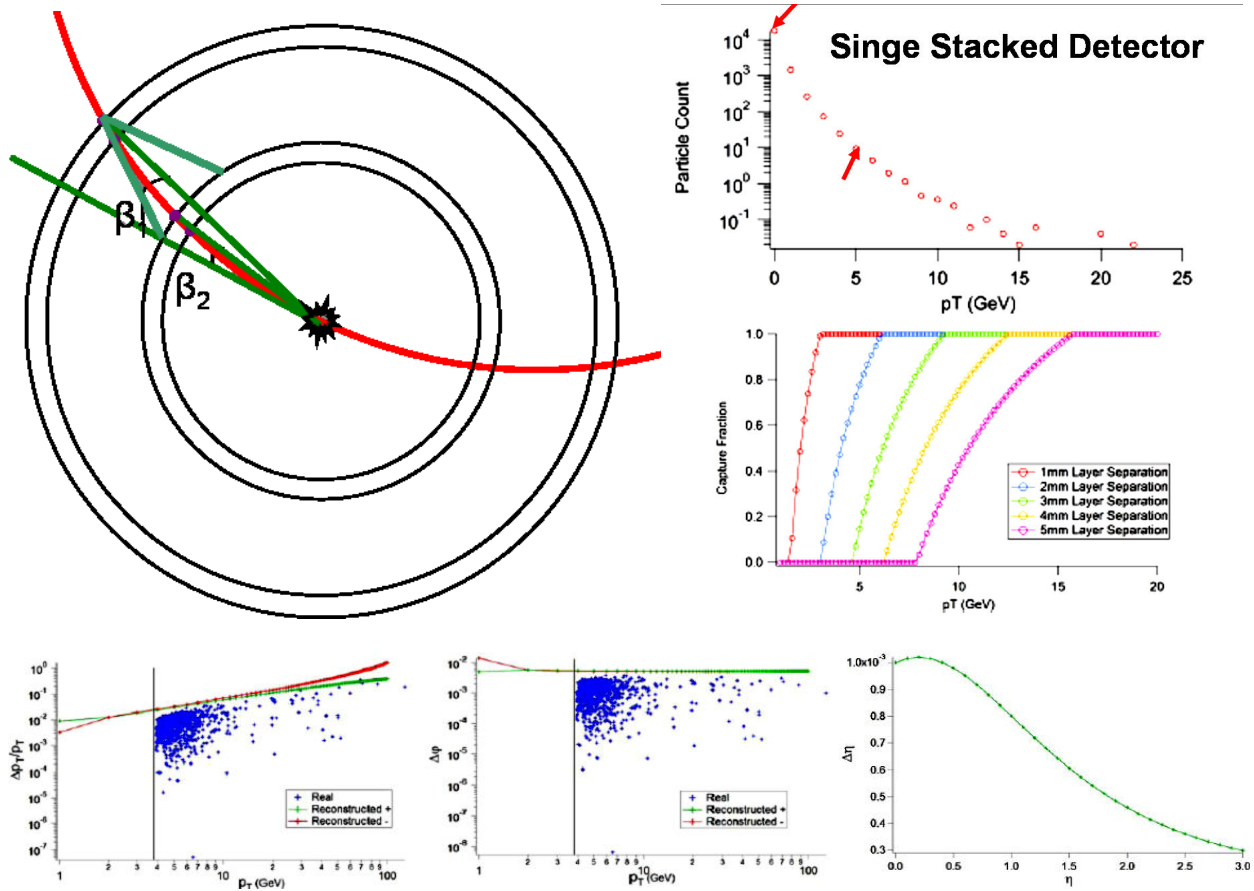

Figure 4: A double stacked pixel detector (top left) for triggering at the SLHC. Charged particle distribution versus transverse momentum (top right). The arrows indicate the reduction achieved by a single stacked detector with a 5 GeV transverse momentum cut. Acceptance versus transverse momentum as a function of the stack radial distance (middle right). Momentum, phi and pseudorapidity resolutions for a double stacked detector as in top left.

Hence, a mechanism is required to reduce the data on the detector and several ideas have been proposed [4,5,6]. The most studied scenario is that of a *stacked pixel* detector [4,5] shown in Fig. 4. The key point is that the large data rate from a pixel detector used as a triggering device comes from low momentum tracks which if rejected would not have a detrimental effect on the trigger efficiency. The stacked detector which has been proposed has the following general characteristics:

---

[3] The numbers given here have been obtained by scaling by 4 the 80 MHz results from [3].

- Two layers of pixellated detectors instrumented with binary readout at radius r = 20 cm.
- A pixel z-ϕ pitch of the order of 50-100 μm.
- Two pixel layers placed a few mm apart radially.
- On-detector electronics form coincidences between pixels of the two layers (trigger stubs) and only stubs with sufficient transverse angle are transmitted off-detector to the trigger primitive generator.

The results from fast Monte Carlo studies [4,5] are shown also in Fig. 4 and demonstrate that:

- The transverse angle cut is equivalent to a $p_T$-cut of sufficient resolution. As seen in Fig. 4 (middle-right) by adjusting the radial distance between the stack layers and the pixel size one can tune the momentum cut.
- Transmitting to the off-detector electronics only hits which correspond to tracks of $p_T$ greater approximately 5 GeV would reduce the rate by at least 3 orders of magnitude (Fig. 4 top-left). Further reduction of 1-2 orders of magnitude can be achieved if the threshold can be increased to 10 GeV.
- Track stubs found this way have sufficient pseudorapidity and azimuthal angle resolution to correlate them with objects found by the calorimeter and muon triggers.
- Including a second double stacked detector at ~10 cm radial distance will significantly increase the $p_T$, pseudorapidity and azimuthal angle resolution of the stubs and could make possible further reductions in the first level trigger rate (Fig. 4 bottom).

## 4. SUMMARY

The CMS trigger has been commissioned and is ready for data taking in the next LHC run. Research and development effort has already started for designing the new Lvl-1 trigger system for CMS at the SLHC. The focus of this effort is the development of a tracking trigger using a stacked tracking detector which will be placed in the inner tracker region.
.